\documentclass[journal=mamobx]{achemso}
\usepackage{graphicx}
\usepackage{caption}
\usepackage{subcaption}
\usepackage{xcolor}
\usepackage[hidelinks]{hyperref}
\usepackage{multirow}
\usepackage{lipsum}
\usepackage{palatino}
\usepackage{amsmath}
\usepackage{amssymb}
\setkeys{acs}{articletitle = true}

\usepackage[normalem]{ulem}

\newif\ifHighlitedChanges
\def\ifHighlitedChanges{\iftrue}
\ifHighlitedChanges
  
  \def\STRIKE#1{{\color{red}\sout{#1}}}
\else
  
  \def\STRIKE#1{\relax}
\fi

\title{Coupling of nanoparticle dynamics to polymer center-of-mass motion in semidilute polymer solutions}

\author{Renjie Chen}
\affiliation{Department of Chemical and Biomolecular Engineering, University of Houston, Houston, TX 77204}
\altaffiliation{Contributed equally to this work}

\author{Ryan Poling-Skutvik}
\affiliation{Department of Chemical and Biomolecular Engineering, University of Houston, Houston, TX 77204}
\altaffiliation{Contributed equally to this work}

\author{Arash Nikoubashman}
\affiliation{Institute of Physics, Johannes Gutenberg University Mainz, Staudingerweg 7, 55128 Mainz, Germany}

\author{Michael P. Howard}
\affiliation{Department of Chemical and Biological Engineering, Princeton University, Princeton, New Jersey 08544}

\author{Jacinta C. Conrad}
\affiliation{Department of Chemical and Biomolecular Engineering, University of Houston, Houston, TX 77204}

\author{Jeremy C. Palmer}
\email{jcpalmer@uh.edu}
\affiliation{Department of Chemical and Biomolecular Engineering, University of Houston, Houston, TX 77204}

\date{\today}

\begin{document}

\begin{abstract}

We investigate the dynamics of nanoparticles in semidilute polymer solutions when the nanoparticles are comparably sized to the polymer coils using explicit- and implicit-solvent simulation methods. The nanoparticle dynamics are subdiffusive on short time scales before transitioning to diffusive motion on long time scales. The long-time diffusivities scale according to theoretical predictions based on full dynamic coupling to the polymer segmental relaxations. In agreement with our recent experiments, however, we observe that the nanoparticle subdiffusive exponents are significantly larger than predicted by the coupling theory over a broad range of polymer concentrations. We attribute this discrepancy in the subdiffusive regime to the presence of an additional coupling mechanism between the nanoparticle dynamics and the polymer center-of-mass motion, which differs from the polymer relaxations that control the long-time diffusion. This coupling is retained even in the absence of many-body hydrodynamic interactions when the long-time dynamics of the colloids and polymers are matched.\end{abstract}

\maketitle
\section{Introduction}
The Brownian dynamics of colloids suspended in a purely viscous fluid is traditionally described by the Stokes-Einstein (SE) equation, which relates the diffusivity $D$ to the ratio of thermal energy of the colloid to the viscous drag over the particle surface. In complex fluids, {\it e.g.} polymer solutions, the colloid dynamics are in addition affected by viscoelastic contributions, which can be incorporated into the generalized Stokes-Einstein (GSE) expression through a complex viscosity $\tilde{\eta}$.\cite{Mason1995, Squires2010} An underlying assumption of both the SE and GSE expressions is that the fluid can be regarded as an effective continuum over the particle surface. When this continuum approximation is broken by a particle that is comparably sized to a characteristic length scale of the material, however, the particle dynamics deviates from these expressions and a description based solely on the zero-shear viscosity of the material becomes insufficient.\cite{Mackay2003, Tuteja2007, Ye1998, Cheng2002,Guo2012,Choi2015,Maldonado-Camargo2016,Pryamitsyn2016} 

Semidilute polymer solutions are non-Newtonian fluids used commonly during polymer composite processing,\cite{Sarkar2015} as sweep fluids in enhanced oil recovery methods,\cite{Wever2011} and to produce hydrogels.\cite{Schultz2009} They serve as ideal models for complex heterogeneous materials because their characteristic length scales, such as the polymer radius of gyration $R_\mathrm{g}$ and the correlation length $\xi$, are well-defined and easily tuned by changing the molecular weight and concentration of the polymer.\cite{Rubinstein2003} Transport of particles or molecules through polymer solutions has traditionally been explained using geometric obstruction models,\cite{Ogston1973, Johansson1991} in which the diffusivity decreases with an increase in polymer concentration due to a higher frequency of collisions between the particles and polymer chains, or using hydrodynamic models,\cite{Cukier1984, Phillies1985, Phillies1986, Cheng2002} in which the polymer chains increase the solution viscosity and screen hydrodynamic interactions. The success of these models, however, is typically limited to a narrow range of particle sizes, polymer molecular weights, or polymer concentrations; furthermore, these models do not specifically address the deviations of nanoparticle dynamics from SE predictions. Recent theoretical treatments have employed modified mode-coupling theory\cite{Egorov2011} or self-consistent Langevin equations\cite{Yamamoto2015} to relate the nanoparticle dynamics to local fluctuations in the polymer mesh. Such treatments accurately reproduce the long-time dynamics of the nanoparticles but have largely not investigated dynamics on shorter time and length scales due to the sensitivity of the analytical calculations to the dynamic propagator of the polymer fluctuations. To model nanoparticle dynamics over a wide range of time and length scales, coupling theory\cite{Cai2011} proposes that the nanoparticle dynamics directly couple to the segmental relaxations of the surrounding polymer chains. Under this assumption, nanoparticles are locally trapped by the polymer chains, leading to subdiffusive motion on short time scales. As the polymer chains relax over the particle surface, the nanoparticle can break out of its local cage and begin to freely diffuse through the solution with a size-dependent diffusivity, which scales as $\sigma_\mathrm{NP}/\xi$, where $\sigma_\mathrm{NP}$ is the nanoparticle diameter.

In our previous experimental work, we found excellent agreement between coupling theory and long-time particle diffusivities, and for the short-time particle dynamics in the limit of small or large nanoparticles relative to the characteristic length scales of the polymer solution (\latin{i.e.}, $\sigma_\mathrm{NP} < \xi$ and  $\sigma_\mathrm{NP}  \gtrsim 10 \xi$, respectively).\cite{Poling-Skutvik2015} Substantial deviations from the predicted behavior, however, were observed at short times for particles of size comparable to the correlation length. Across a broad range of polymer concentrations, the nanoparticle sub\-diffusive exponents $\alpha_\mathrm{NP}$ were much larger than predicted, and varied with both particle size and polymer concentration. Additionally, we found that long-range interparticle interactions affected the sub\-diffusive motion of the particles.\cite{Poling-Skutvik2016} This result suggests that the energy barrier for particle motion through the polymer mesh on short time and length scales is finite, in contrast to the infinite barrier required for full coupling of the particle and polymer dynamics. Thus, despite the notable success of the widely-applied coupling theory in describing the long-time dynamics in experiments\cite{Omari2009,Poling-Skutvik2015} and simulations,\cite{liu2008molecular} the physics underlying the subdiffusive particle dynamics on short time scales in polymer solutions remains incompletely understood. Critical open questions are what causes the short-time subdiffusive dynamics and what controls the crossover to long-time diffusion. This understanding is essential for predicting particle transport and dispersion during composite \cite{Sarkar2015} and hydrogel\cite{Schultz2009} processing and in oil production and exploration.\cite{Wever2011} Simulations are an ideal method to probe short-time dynamics and have been extensively used to investigate nanoparticle dynamics in polymer melts.\cite{starr2002molecular, brown2003molecular, liu2008molecular} Extending these methods to investigate dynamics in polymer solutions, however, remains challenging due to the computationally demanding nature of accurately modeling solvent-mediated interactions. These interactions are strongly screened in melts, but influence short-time dynamics in polymer solutions.

Here, we simulate the dynamics of nanoparticles in semidilute solutions of comparably sized polymers, using multiparticle collision dynamics (MPCD) to account for solvent-mediated hydrodynamic interactions (HI).\cite{Malevanets1999, tao2008multiparticle, Gompper2009, kapral2008multiparticle, Huang2010,Jiang2013, Nikoubashman2014, Howard2015, Nikoubashman2016, Nikoubashman2017a, Nikoubashman2017b} Complementary Langevin dynamics (LD) simulations, which remove HI between particles, are also performed. The friction coefficients employed in the LD simulations are chosen to reproduce the long-time nanoparticle and polymer center-of-mass diffusion coefficients calculated from the MPCD simulations in the dilute regime, allowing us to study short and intermediate time dynamics in the absence of HI while approximately preserving the long-time relaxation behavior observed in the MPCD simulations. The MPCD simulations reveal trends that are qualitatively similar to previous experiments -- the nanoparticle dynamics are subdiffusive on short time scales and diffusive on long time scales, with diffusivities that scale as predicted by coupling theory. The subdiffusive exponents, however, are larger than the predicted value of 0.5 and smoothly decrease with increasing polymer concentration, similarly to what is observed experimentally. Moreover, the subdiffusive exponents of the particle are strongly correlated to those of the polymer center-of-mass, suggesting that coupling to the motion of the polymer center-of-mass may provide an additional mechanism by which the nanoparticles can move through a polymer solution. This coupling mechanism appears in both MPCD and LD simulations, indicating that many-body hydrodynamic interactions are not required when the long-time dynamics are preserved.

\section{Model and Simulation Methods}

Following recent studies,\cite{Huang2010, Nikoubashman2014, Howard2015, Nikoubashman2016, Nikoubashman2017a} we model the polymers in solution as bead-spring chains composed of $N_\text{m}$ monomer beads with diameter $\sigma_{\text{P}}$. Polymer bonds are described by the finitely extensible nonlinear elastic (FENE) potential:\cite{bishop1979molecular}
\begin{equation}
 U_\text{FENE}(r) =
  \begin{cases}
    -\frac{1}{2}\kappa{r_0}^2 \ln\left[1-\dfrac{r^2}{{r_0}^2}\right], & r\leq{r_0} \\
    \infty, & r>r_0 
  \end{cases},
\end{equation}
where $r$ is the scalar separation distance between two bonded beads. Excluded volume interactions are modeled using the purely repulsive, shifted Weeks-Chandler-Andersen (sWCA) potential to simulate good solvent conditions:\cite{weeks1971role}
\begin{equation}
  U_\text{sWCA}(r) =
  \begin{cases}
    4\varepsilon\left[\left(\dfrac{\sigma_{ij}}{r-\Delta_{ij}}\right)^{12}-\left(\dfrac{\sigma_{ij}}{r-\Delta_{ij}}\right)^6\right] + \varepsilon, & r\leq 2^{1/6}\sigma_{ij} +\Delta_{ij} \\
    0, & r>2^{1/6}\sigma_{ij}+\Delta_{ij}
  \end{cases},
\end{equation}
where $\varepsilon$ controls the strength of the repulsion.  For monomer-monomer interactions, we set $\Delta_{ij}=0$ and $\sigma_{ij} = \sigma_{\text{P}}$.  Similarly, for nanoparticle-nanoparticle interactions, $\Delta_{ij}=0$ and $\sigma_{ij} = \sigma_{\text{NP}}$.  To account for the size asymmetry of the polymer monomers and nanoparticles, however, we use $\Delta_{ij}=(\sigma_\text{NP}-\sigma_\text{P})/2$ and $\sigma_{ij}=\sigma_{\text{P}}$ for the cross interactions. 

\begin{figure}[!htbp]
\includegraphics[width=2.8 in]{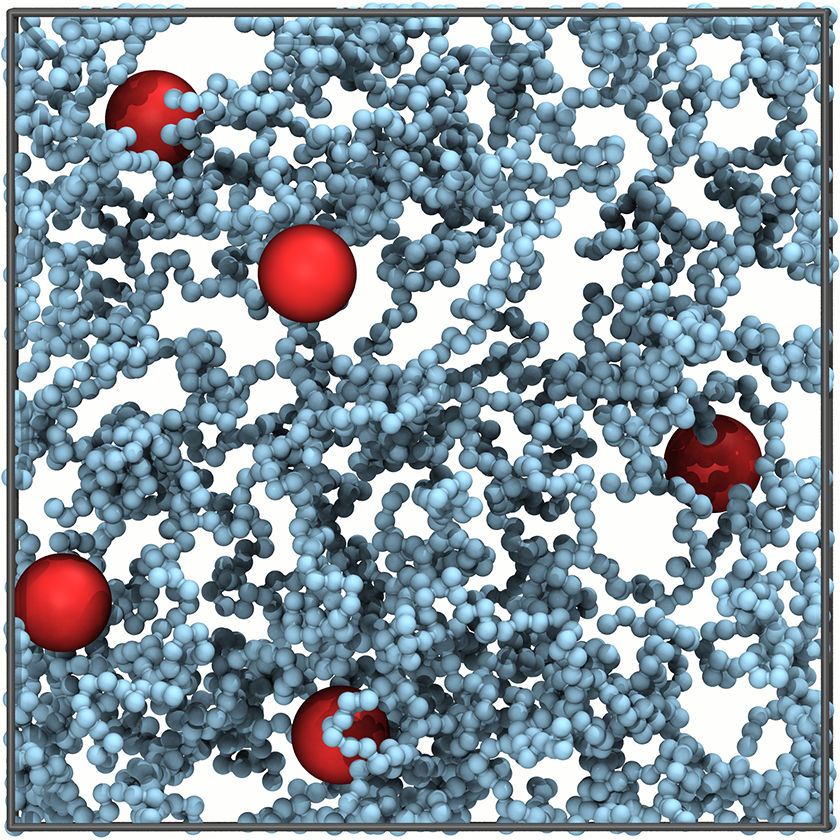}
\caption{\label{fig:simulation_rendering} Rendering of nanoparticles (red) dispersed in a solution of polymers (blue) at $c/c^*=0.5$  simulated in our study.}
\end{figure}

All simulations were performed in a cubic box with a $40\sigma_{P}$ edge length and periodic boundary conditions in all directions. We used LAMMPS\cite{plimpton1995fast} to conduct our simulations. A value of $\varepsilon = k_{\rm B} T $ was used for all particle interactions, where $k_{\rm B}$ is Boltzmann's constant and $T$ is temperature. The polymers were modeled using $N_\text{m} =50$ beads and the standard Kremer-Grest parameters  for the bonded interactions,\cite{grest1986molecular} \latin{i.e.} $\kappa=30k_{\rm B}T\sigma_{P}^{-2}$ and $r_0=1.5\sigma_{P}$. Simulations using a larger box size ($60\sigma_{P}$) or longer polymer chains (up to $N_{\rm m} = 250$) revealed no qualitative differences in the scaling behavior of the nanoparticle dynamics. The radius of gyration at infinite dilution for this model is $R_{g,0}=4.9 \sigma_{\text{P}}$, leading to an overlap concentration of $c^* = N_{\text{m} }\times \left(4 \pi R_{g,0}^3/3\right)^{-1} = 0.1 \sigma_{\text{P}}^{-3}$. Hence, for the range of concentrations investigated here ($0.2 - 8.0 c/c^*$), the simulated systems contain between 25 and 1000 polymer chains.  For the nanoparticles, we set $d_\text{NP} = 6 \sigma_{\text{P}}$, such that they are similar in size to the polymer coils. Figure~\ref{fig:simulation_rendering} shows a typical simulation snapshot for $c/c^*=0.5$.

Many-body HI in the nanoparticle-polymer solutions were simulated using the MPCD algorithm.\cite{Malevanets1999, tao2008multiparticle, Gompper2009, kapral2008multiparticle} In MPCD, mesoparticles (polymers and nanoparticles) are immersed in an background solvent, which is modeled explicitly through an ensemble of point particles. These solvent particles exchange momentum with nearby solvent and mesoparticles through stochastic collisions, which are designed to ensure that hydrodynamic correlations emerge over sufficiently large length scales.\cite{Gompper2009} The MPCD simulations for our model were conducted using a momentum conserving version of the Andersen thermostat \cite{allahyarov2002mesoscopic, noguchi2007particle} that we implemented into the existing stochastic rotation dynamics (SRD) module in LAMMPS.\cite{bolintineanu2014particle} This scheme, which is often referred to as MPCD-AT, is described in detail elsewhere.\cite{allahyarov2002mesoscopic, noguchi2007particle, Theers2015} The MPCD routines for LAMMPS used in our study are available online \bibnote{MPCD routines and sample input files for LAMMPS are available on GitHub: \href{https://github.com/palmergroup/mpcd_polymer_colloid}{\url{https://github.com/palmergroup/mpcd_polymer_colloid}}}, along with example scripts for simulating solutions of polymers and nanoparticles.

The edge length, $a$, of the cubic MPCD collision cells dictates the spatial resolution of the HI,\cite{Huang2012} and we chose $a = \sigma_{\text{P}}$ for our simulations. We assigned unit mass $m=1$ to each solvent particle and used an average MPCD solvent density $\rho = 5 m/\sigma_\text{P}^3$. The collision time step was set to $\Delta t = 0.09 \tau$, where $\tau =\sqrt{m\sigma^2_{\text{P}} /\left(k_BT\right)}$ is the unit of time in the simulations. The reference positions of the cell were also randomly shifted before each collision step to ensure Galilean invariance.\cite{ihle2001stochastic} These typical parameters give rise to an MPCD solvent with Schmidt number $\mathrm{Sc} \approx 12.0$ and dynamic viscosity $\eta_0 \approx 4.0 \tau k_{\rm B} T /\sigma_{\text{p}}^3$. The motions of the polymers and nanoparticles in the MPCD simulations were integrated using a velocity-Verlet scheme with a $0.002 \tau$ time step.  Momentum transfer between solvent particles and polymers during the collision steps was handled using the scheme described in Ref.\ \citenum{Huang2010}, whereas solvent collisions with the nanoparticles were treated using the stochastic boundary algorithm discussed in Ref.\ \citenum{bolintineanu2014particle} with slip conditions. The masses of the monomers ($M_\text{P} = \rho \sigma_\text{ P}^3$) and nanoparticles ($M_\text{NP} = \rho \pi \sigma_\text{NP}^3 / 6$) were set to achieve neutral buoyancy in the background solvent. Dynamic properties reported for the nanoparticles were obtained by averaging over at least 30 trajectories.

We also performed a complementary set of LD simulations to investigate the behavior of the solutions in the absence of many-body HI. The LD friction coefficients\cite{turq1977brownian} for the nanoparticles and polymer monomers were adjusted independently to match long-time diffusion behavior from MPCD simulations of dilute solutions. The nanoparticle friction coefficient was chosen to reproduce the long-time nanoparticle diffusivity in pure MPCD solvent. Similarly, the friction coefficient for the monomers was chosen in such a way that the long-time polymer center-of-mass diffusion coefficient from MPCD and LD simulations matched in the absence of nanoparticles at $c/c^*=0.2$.
   
\section{Results and Discussion}
\begin{figure}[!htbp]
\includegraphics{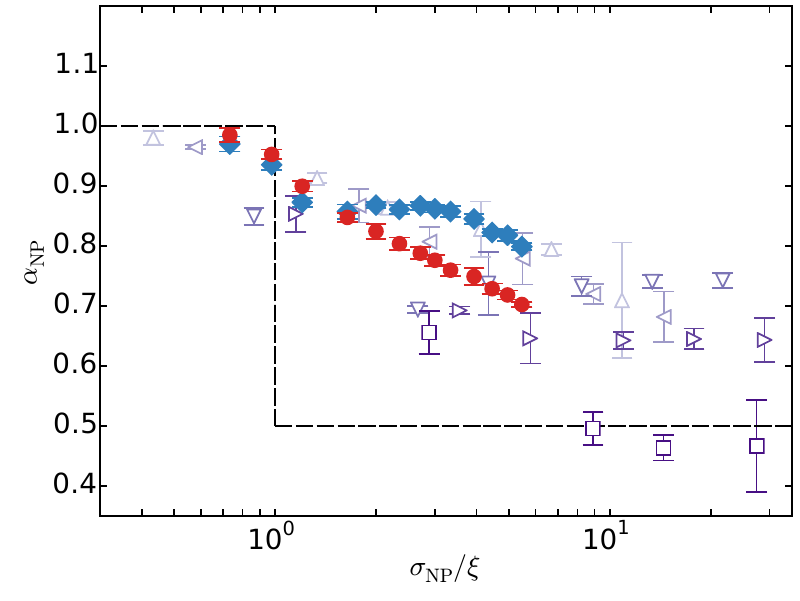}
\caption{\label{fig:AlphaLength} Nanoparticle subdiffusive exponent $\alpha_\mathrm{NP}$ as a function of the ratio of nanoparticle diameter $\sigma_\mathrm{NP}$ to correlation length $\xi$ for MPCD (red circles) and LD (blue diamonds) simulations with $\sigma_\mathrm{NP}/2R_\mathrm{g,0} = 0.61$. Open symbols are experimental data from Ref.\ \citenum{Poling-Skutvik2015} for particles with $\sigma_\mathrm{NP}/2R_\mathrm{g,0} = 0.56$ ($\vartriangle$), 0.74 ($\triangleleft$), 1.1 ($\triangledown$), 1.5 ($\triangleright$), and 3.7 ($\square$). Dashed line is prediction from coupling theory in Ref.\ \citenum{Cai2011}.  For  $c/c^* > 1$, $\xi = R_\mathrm{g,0} (c/c^*)^{-\nu/(3\nu-1)}$ with $\nu \approx$ 0.62 (experiment) and 0.61 (simulation).  For $c/c^* \le 1$, we calculate $\xi$ according to the mean geometric separation distance $R_\mathrm{g,0} (c/c^*)^{-1/3}$.}
\end{figure}

Coupling theory\cite{Cai2011} predicts that the nanoparticle dynamics are subdiffusive on short time scales with a mean-squared displacement (MSD) that scales as a power-law in time, $\left< \Delta r^2 \right> \sim t^{\alpha_\mathrm{NP}}$. Nanoparticles smaller than the polymer correlation length (\latin{e.g.} $\sigma_\mathrm{NP} < \xi$) are predicted to pass freely through the polymer mesh, so that $\alpha_\mathrm{NP} = 1$. Once $\sigma_\mathrm{NP} > \xi$, the particles are predicted to be locally trapped by the polymer and can only move according to the segmental Rouse relaxations of the surrounding chains, so that $\alpha_\mathrm{NP} = 0.5$. Indeed, this sharp transition has been observed in previous experiments where the dispersed nanoparticles were chemically bound to transient polymer networks.\cite{Sprakel2007} In our MPCD simulations of athermal nanoparticle-polymer solutions, however, we observe a smooth, monotonic decay in $\alpha_\mathrm{NP}$ rather than the predicted step function from $\alpha_\mathrm{NP} = 1$ to 0.5 at $\sigma_\mathrm{NP}/\xi = 1$ (Fig.\ \ref{fig:AlphaLength}).  A similar trend was observed in our previous experiments,\cite{Poling-Skutvik2015} recovering $\alpha_\mathrm{NP}=0.5$ only for large particles at high polymer concentrations. The MPCD simulations and experiments therefore collectively demonstrate that coupling theory correctly captures the dynamic behavior in the limits $\sigma_\mathrm{NP} < \xi$ and $\sigma_\mathrm{NP} \gtrsim 10\xi$. However, they also reveal a surprisingly broad crossover regime, where the subdiffusive exponent is significantly larger than expected from coupling theory.

Comparison of the LD and MPCD simulations also suggests that HI influence the short-time nanoparticle dynamics. In the dilute limit (i.e.  $c/c^* \rightarrow 0$), both LD and MPCD predict that $\alpha_\mathrm{NP} \rightarrow 1$, in accord with experiment (Fig.\ \ref{fig:AlphaLength}).  This behavior indicates that the nanoparticle dynamics become purely diffusive after transitioning from a ballistic regime on much shorter time scales. Agreement between the two simulation methods is expected in this regime because the LD friction coefficients are matched to explicitly reproduce the long-time diffusive relaxations from MPCD.  Progressively larger deviations between LD and MPCD are observed, however, as $\sigma_\mathrm{NP} / \xi$ (and $c/c^*$) increase.  For $1 \lesssim \sigma_\mathrm{NP}/\xi\lesssim 4$, $\alpha_\mathrm{NP}$ in LD is approximately constant and deviates from the steady decay with concentration observed in the MPCD simulations, suggesting that many-body HI strongly affect the dynamics in this regime. Finally, for $\sigma_\mathrm{NP}/\xi  \gtrsim 4$, $\alpha_\mathrm{NP}$ from LD and MPCD scale similarly with polymer concentration. This behavior is consistent with the expectation that many-body HI are screened in concentrated polymer solutions.\cite{DeGennes1979, Richter1984, Huang2010,Ahlrichs2001}

\begin{figure*}[!htbp]
\includegraphics{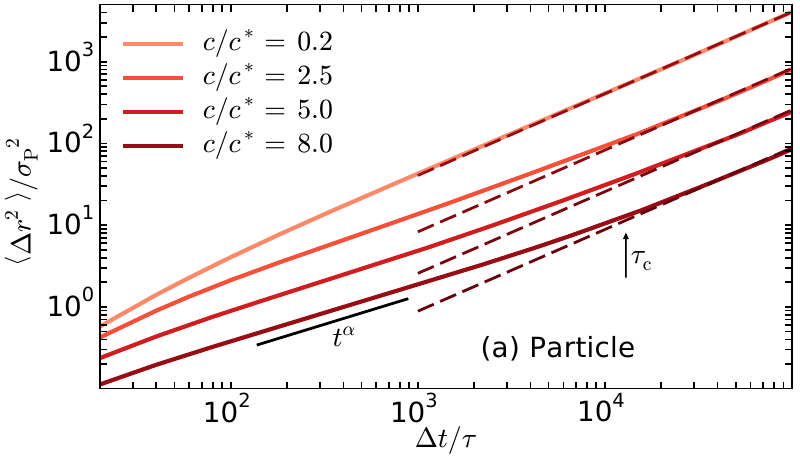} 
\hspace{10 pt}
\includegraphics{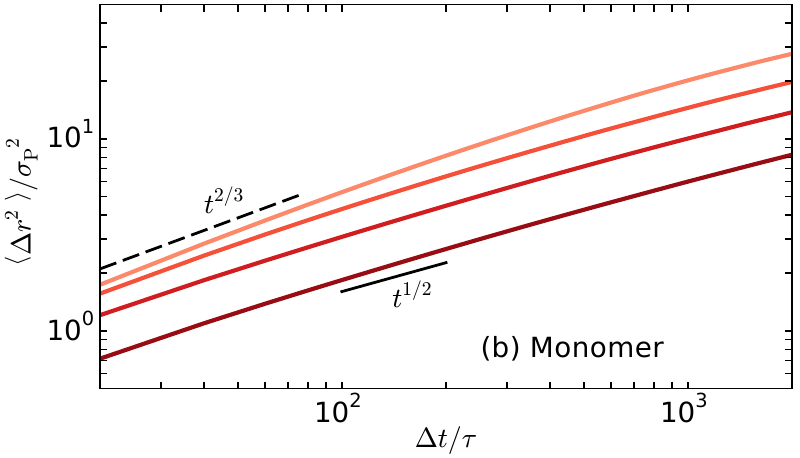}
\hspace{10 pt}
\includegraphics{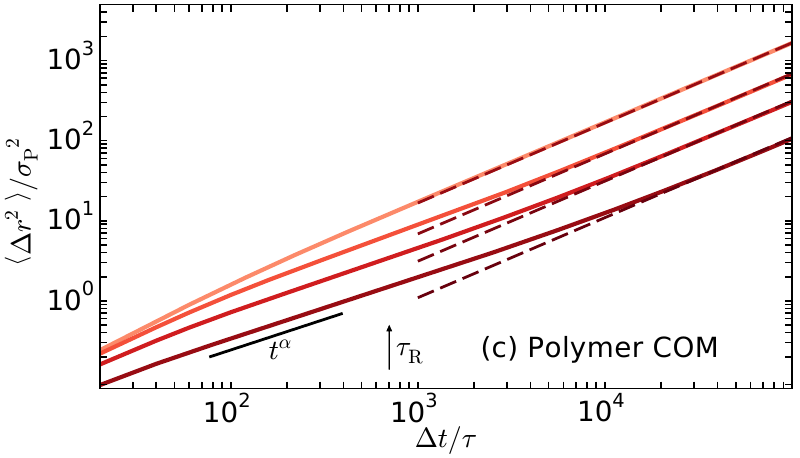}
\caption{\label{fig:msds} Mean-squared displacement $\left< \Delta r^2 \right>$ extracted from MPCD simulations normalized by bead diameter $\sigma_\mathrm{P}$ as a function of normalized lag time $\Delta t /\tau$ at multiple polymer concentrations for (a) nanoparticles, (b) monomer beads in the polymer center-of-mass reference frame, and (c) the polymer center-of-mass (COM). Dashed and solid lines in (a) and (c) indicate diffusive and subdiffusive dynamics, respectively. Dashed and solid lines in (b) indicate predicted scaling for the Zimm and Rouse modes, respectively. Arrows indicate representative time scales for the $8c^*$ solution, including the observed diffusive crossover time $\tau_\mathrm{c}$ and estimated Rouse relaxation time  $\tau_\mathrm{R} = \tau_\xi (R_\mathrm{g}/\xi)^4$. The Zimm time $\tau_\xi \approx \eta_0 \xi^3/k_\mathrm{B}T$ is estimated to be $\tau_\xi \sim 10 \tau$ for $8c^*$ and hence is not shown.}
\end{figure*}

To investigate these short-time dynamics, it is instructive to examine the behavior of both nanoparticles and polymer chains in the MPCD simulations. Qualitatively, the MSD $\left< \Delta r^2 \right>$ of the nanoparticles (Fig.\ \ref{fig:msds}(a)) exhibits the predicted features \cite{Cai2011} that have been observed in experiments \cite{BabayeKhorasani2014, Poling-Skutvik2015} and simulations. \cite{li2016diffusion} On short time scales, the particles move subdiffusively with $\alpha_\mathrm{NP} < 1$. On long time scales, the particle motion becomes diffusive (\latin{i.e.} $\alpha_\mathrm{NP} = 1$) with a diffusivity $D$ that decreases with increasing polymer concentration. As an additional verification of the simulations, the MSD for the monomer beads in the center-of-mass reference frame (Fig.\ \ref{fig:msds}(b)) exhibits the expected Zimm scaling \big($\left< \Delta r^2 \right> \sim t^{2/3}$\big) at low polymer concentrations and Rouse scaling \big($\left< \Delta r^2 \right> \sim t^{1/2}$\big) at higher polymer concentrations. The transition from Zimm to Rouse relaxations confirms that HI are screened at high polymer concentrations, in agreement with polymer scaling predictions.\cite{DeGennes1979} Finally, the MSD for the polymer chain center-of-mass (Fig.\ \ref{fig:msds}(c)) exhibits qualitatively similar behavior to that of the nanoparticles. On short time scales, the polymer center-of-mass motion is subdiffusive with an exponent $\alpha_\mathrm{P} < 1$, similar to what has been observed previously in molecular dynamics simulations\cite{Padding2001} and experiments\cite{Smith2001} for polymer chains in unentangled melts. On long time scales, the polymer relaxations are dominated by the longest Rouse mode, so that the center-of-mass moves diffusively. Additionally, comparison of the nanoparticle and polymer center-of-mass MSDs at the same polymer concentration indicates that both nanoparticles and polymer chains are mobile over similar time and length scales. 

\begin{figure}[!htbp]
\includegraphics{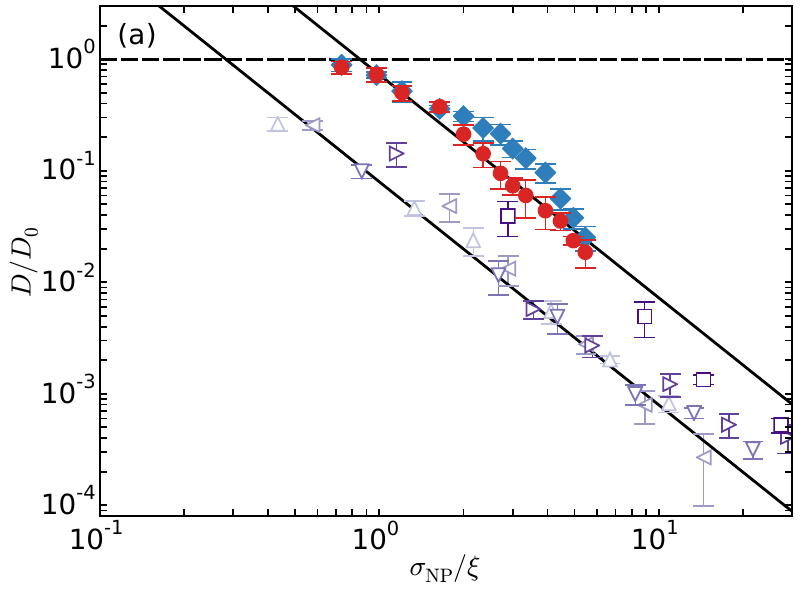}
\hspace{10 pt}
\includegraphics{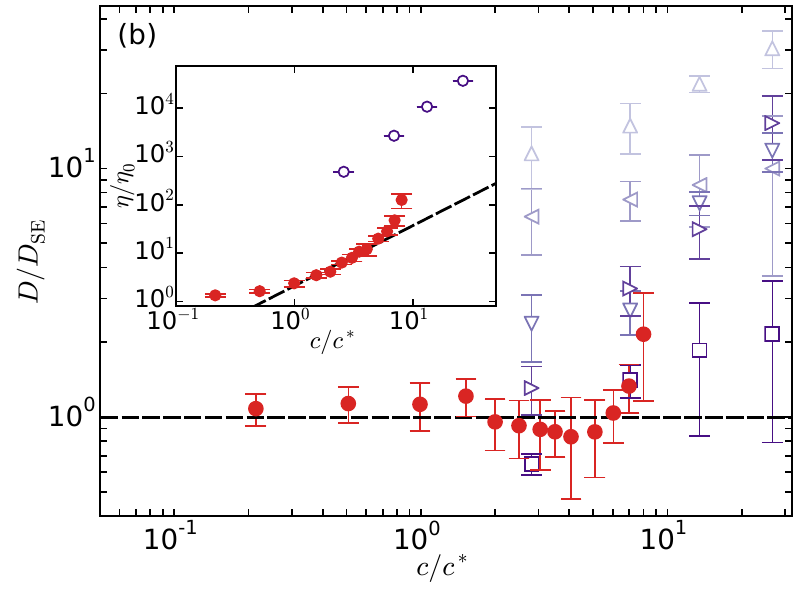}
\caption{\label{fig:NormDiff} (a) Normalized nanoparticle diffusivity $D/D_0$ as a function of normalized size $\sigma_\mathrm{NP}/\xi$ in MPCD (red circles) and LD (blue diamonds) simulations with $\sigma_\mathrm{NP}/2R_\mathrm{g,0} = 0.61$. Open symbols are experimental data from Ref.\ \citenum{Poling-Skutvik2015} for particles with $\sigma_\mathrm{NP}/2R_\mathrm{g,0} = 0.56$ ($\vartriangle$), 0.74 ($\triangleleft$), 1.1 ($\triangledown$), 1.5 ($\triangleright$), and 3.7 ($\square$). Solid lines show coupling theory predictions from Ref.\ \citenum{Cai2011}.  (b) Ratio of nanoparticle diffusivity to Stokes-Einstein prediction $D_\mathrm{SE}$ calculated from zero-shear solution viscosity as a function of polymer concentration $c/c^*$. \textit{Inset}: Zero-shear viscosity of MPCD (closed symbols) and experimental solutions (open symbols) as a function of polymer concentration. Dashed line indicates predicted scaling behavior for an ideal polymer solution. For  $c/c^* > 1$, $\xi = R_\mathrm{g,0} (c/c^*)^{-\nu/(3\nu-1)}$ with $\nu \approx$ 0.62 (experiment) and 0.61 (simulation).  For $c/c^* \le 1$,  we calculate $\xi$ according to the mean geometric separation distance $R_\mathrm{g,0} (c/c^*)^{-1/3}$.}
\end{figure}

Confirming that the MPCD simulations accurately capture the polymer relaxations, we now analyze the change in long-time particle diffusivity with increasing polymer concentration (Fig.\ \ref{fig:NormDiff}). At low polymer concentrations where $\sigma_\mathrm{NP}/\xi < 1$, the particle diffusivities are similar to those observed in pure solvent so that $D/D_0 \approx 1$, where $D_0 \sim k_\mathrm{B}T/\eta_0 \sigma_\mathrm{NP}$ is the diffusivity of the particle in the absence of polymer and $\eta_0$ is the solvent viscosity. For $\sigma_\mathrm{NP}/\xi > 1$, the particle diffusivities scale as $D/D_0 \sim (\sigma_\mathrm{NP}/\xi)^{-2}$, as predicted from coupling theory,\cite{Cai2011} where $\xi = R_\mathrm{g,0} (c/c^*)^{-\nu/(3\nu-1)}$ is the polymer correlation length \cite{Rubinstein2003} and $\nu$ is the inverse of the polymer fractal dimension.   For the bead-spring polymer model considered here $\nu=0.61$, which is in good agreement with previous computational studies\cite{Huang2010}, with the estimated value of 0.62 for the partially hydrolyzed polyacrylamide used in experiment,\cite{Poling-Skutvik2015} and with the theoretically predicted value of 0.59 for flexible chains in a good solvent\cite{Huang2010}. 

The scaling of the nanoparticle diffusivities with $\sigma_\mathrm{NP}/\xi$ for MPCD simulations agrees with that observed in our recent experiments\cite{Poling-Skutvik2015}. The offset between the simulated and experimental data is attributable in large part to the difference in shear viscosities (inset to Fig.\ \ref{fig:NormDiff}(b)). In both experiments and simulations, the solution viscosity $\eta$ was determined through shear measurements in the linear response regime\cite{MullerPlathe1999}. Whereas the simulations use a generic monodisperse polymer model in a good solvent, the polymers used in experiment were highly polydisperse polyelectrolytes with a charge functionality of $\approx 30$\%. It is well established that the viscosity of solutions of charged polymers scales differently with concentration than that of neutral polymer solutions,\cite{Dobrynin2005,Colby2009} resulting in the order of magnitude difference between the viscosities of the experimental and simulated solutions (inset to Fig.\ \ref{fig:NormDiff}(b)). Specifically, the shear viscosity of the experimental solution is approximately $60$ times higher for $c/c^* = 3$. Hence, the fact that the normalized diffusivities $D/D_0$ for the experiments are lower than those computed from simulation is expected. Nevertheless, both data sets exhibit the same qualitative trends.

To assess the changes in particle diffusivity relative to bulk solution properties, we compare the diffusivities from simulation and experiment to those predicted by the Stokes-Einstein predictions $D_\mathrm{SE} \sim k_\mathrm{B}T/\eta \sigma_\mathrm{NP}$, where $\eta$ is the zero-shear solution viscosity. For $c/c^* \lesssim 5$, the diffusivities obtained from simulations are in statistical agreement with the Stokes-Einstein predictions (Fig.\ \ref{fig:NormDiff}(b)).  At higher concentrations, however,  the diffusivities begin to increasingly deviate from $D_\mathrm{SE}$, in accord with experiments.  Thus, the MPCD simulations accurately capture the change in the long-time nanoparticle dynamics with increasing polymer concentration, and semi-quantitatively describe the deviations from Stokes-Einstein predictions at high concentrations.

Finally, we assess the effects of HI by comparing the diffusivities extracted from the MPCD simulations to those from the LD simulations (Fig.\ \ref{fig:NormDiff}(a)). LD is an implicit-solvent simulation method that captures only the viscous drag on the individual particles but no short- or long-range HI. The friction coefficients for the dispersed nanoparticles and monomers were chosen in such a way to match the long time diffusivity computed from the MPCD simulations under dilute conditions. Hence, as expected, we observe agreement between the diffusivities extracted from LD and from MPCD simulations in the dilute regime for $\sigma_\mathrm{NP}/\xi \lesssim 1$, where the dynamics were matched by construction.  For $\sigma_\mathrm{NP}/\xi \gtrsim 5$, where HI are strongly screened \cite{DeGennes1979, Richter1984, Huang2010,Ahlrichs2001}, the diffusivities from LD and MPCD also exhibit similar scaling behavior. However, deviations between LD and MPCD, similar to those observed for the subdiffusive exponent (Fig.\ \ref{fig:AlphaLength}), occur in the intermediate regime. Thus our results suggest that HI influence the long-time particle diffusivities for $\sigma_\mathrm{NP}/\xi \lesssim 5$.\cite{Chen2017}

\begin{figure}[ht]
\includegraphics{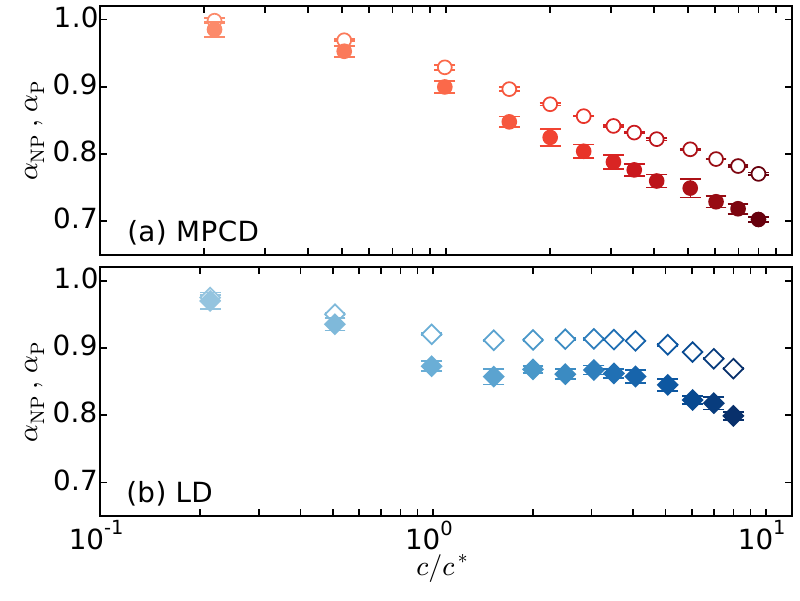}
\hspace{10 pt}
\includegraphics{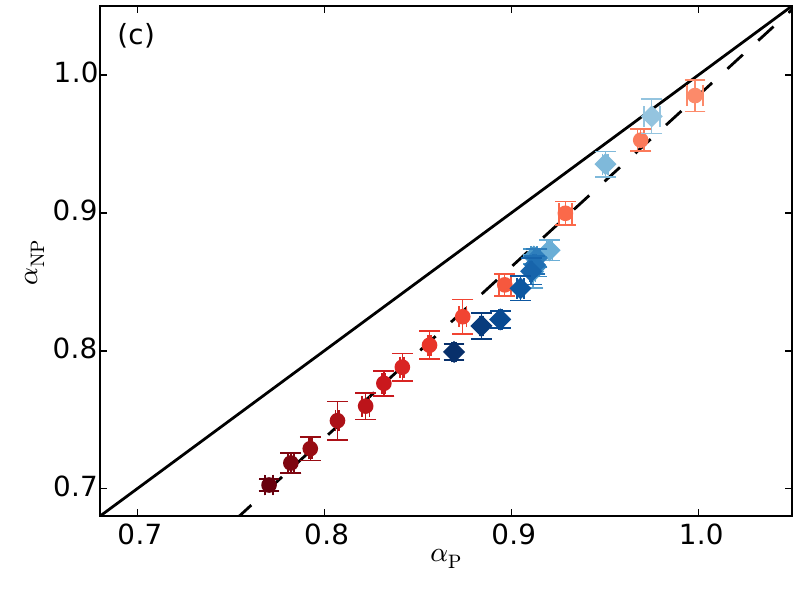}
\caption{\label{fig:AlphaAlpha} Subdiffusive exponents $\alpha_\mathrm{NP}$ (closed) and $\alpha_\mathrm{P}$ (open) for nanoparticle and polymer center-of-mass dynamics, respectively, as a function of polymer concentration in (a) MPCD (red circles) and (b) LD (blue diamonds) simulations. (c) Correlation between $\alpha_\mathrm{NP}$ and $\alpha_\mathrm{P}$.  Solid and dashed lines are the identity line and linear fit to MPCD data, respectively.}
\end{figure}

In addition to the long-time nanoparticle dynamics, the MPCD and LD simulations provide crucial insights into the coupling between nanoparticles and polymer chains on short time and length scales, which are difficult to experimentally measure. Coupling theory assumes that the longest relaxation time of the polymer $\tau_\mathrm{R} = \tau_\xi (R_\mathrm{g}/\xi)^4$ is much larger than the crossover time $\tau_\mathrm{c}$ at which the nanoparticle dynamics transition from subdiffusive to diffusive,\cite{Cai2011} so that the particle dynamics are fully coupled to the polymer segmental relaxations. Under this assumption, nanoparticle dynamics become diffusive once the polymer segments relax over the particle surface. For the simulated polymer chains, the calculated Rouse time ranges from $10^2 \tau$ to $10^3\tau$ depending on polymer concentration, in good agreement with when the polymer center-of-mass begins moving diffusively (Fig.\ \ref{fig:msds}(c)). Whereas coupling theory assumes a separation of time scales, this Rouse time scale is comparable to the crossover time of the particles ($\tau_\mathrm{c} \approx \tau_\mathrm{R}$), indicating that the polymer center-of-mass motion cannot be neglected. Comparable time scales were also observed in our previous experiments \cite{Poling-Skutvik2015}. Based on the similarities of the MSDs (Fig.\ \ref{fig:msds}), we compare the subdiffusive exponents for the nanoparticles and the polymer center-of-mass (Fig.\ \ref{fig:AlphaAlpha}(a,b)). At low polymer concentrations, both particles and polymer chains move diffusively (\latin{i.e.} $\alpha_\mathrm{NP}, \alpha_\mathrm{P} \approx 1$). As the polymer concentration increases, the particles and polymer chains become subdiffusive with monotonically decreasing sub\-diffusive exponents. Furthermore, the change in nanoparticle and polymer sub\-diffusive exponents are similar in magnitude and shape, indicating that the particle and polymer dynamics on short time scales are positively correlated. The sub\-diffusive exponents of the nanoparticles are slightly lower than those of the polymers but are highly correlated over the entire concentration range (Fig.\ \ref{fig:AlphaAlpha}(c)).  Comparison of MPCD and LD simulations reveals differences in the scaling of the sub\-diffusive exponents and long-time diffusion coefficients at low to intermediate polymer concentrations due to HI (Figs.\ \ref{fig:AlphaLength} and \ref{fig:NormDiff}). Nonetheless, both methods find strong correlation between the nanoparticle and polymer center-of-mass sub\-diffusive exponents. This key finding suggests that the coupling is due to the comparable relaxation time scales of the nanoparticles and polymers and not explicitly due to many-body HI.  The high degree of correlation between nanoparticle and polymer dynamics within the sub\-diffusive regime indicates that the center-of-mass polymer motion may indeed play a role in controlling the nanoparticle dynamics. 

\begin{figure}[ht]
\includegraphics{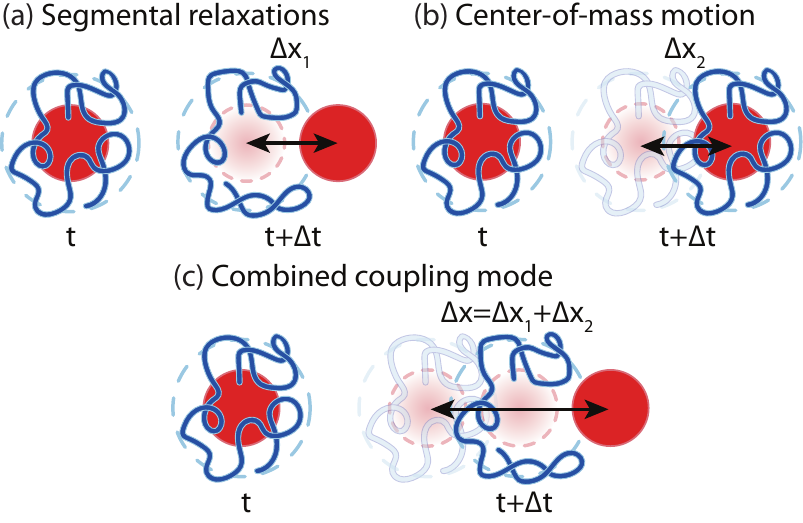}
\caption{\label{fig:picture} Schematic illustrating the physical processes controlling nanoparticle dynamics in semidilute polymer solutions. (a) Segmental relaxation, described in Ref.\ \citenum{Cai2011}. (b) Center-of-mass motion of the polymer. (c) Combined coupling mode includes both segmental relaxation and center-of-mass motion of the polymer.}
\end{figure}

The fact that the short-time nanoparticle dynamics are consistently more subdiffusive than those of the polymer center-of-mass ($\alpha_\mathrm{NP} \lesssim \alpha_\mathrm{P}$, Fig.\ \ref{fig:AlphaAlpha}(c)) suggests that the nanoparticle dynamics may be coupled to additional relaxation modes beyond the polymer center-of-mass dynamics. Because polymer chains are fractal in structure, there is a distribution of relaxation mechanisms that control polymer dynamics. Moving over similar time scales as the polymer, the nanoparticles likely couple to this distribution -- from segmental Rouse motions to center-of-mass diffusion. Indeed, coupling theory\cite{Cai2011} predicts that the segmental motions should play a role in controlling the subdiffusive nanoparticle dynamics. Moreover, from this work and previous experiments, we observe that nanoparticle coupling to segmental relaxations accurately predicts the long-time nanoparticle diffusion across orders of magnitude in polymer concentration and particle size, while the subdiffusive exponents of the nanoparticles and polymer center-of-mass are highly correlated over an order of magnitude in polymer concentration. To combine these contributions into a unified picture, the data suggest that the nanoparticles generally move through polymer solutions via two mechanisms -- coupling to segmental relaxations to move \emph{relative} to the polymer center-of-mass and coupling to the center-of-mass motion to move \emph{with} the polymer center-of-mass (Fig.\ \ref{fig:picture}). The combination of these two mechanisms may lead to the long-time diffusivity of nanoparticles that scales according to the length-scale ratio $\sigma_\mathrm{NP}/\xi$ and deviates from the zero-shear solution viscoelasticity, and to the short-time subdiffusive dynamics with subdiffusive exponents $0.5 \leq \alpha_\mathrm{NP} \lesssim \alpha_\mathrm{P} \leq 1$ and crossover times $\tau_\mathrm{c} \approx \tau_\mathrm{R}$. 

While finalizing our article, Chen et al.\cite{Chen2017} published an MPCD study in which they observed similar subdiffusive behavior for nanoparticles in semidilute polymer solutions.  Our findings are in good agreement with the results from their MPCD simulations performed with HI. Although they used a different interpretive framework based on the empirical model of Ho{\l}yst and collaborators,\cite{Kalwarczyk2011, Kalwarczyk2015} the long-time diffusion coefficients and subdiffusive exponents from their simulations can be collapsed on the same scaling plots shown here and thus support our conclusions. The empirical Ho{\l}yst model \cite{Kalwarczyk2011, Kalwarczyk2015} was not used in this study because, as we previously documented in the Supporting Information of Ref.\ \citenum{Poling-Skutvik2015}, it does not collapse the experimental data as well as the coupling theory of Ref.\ \citenum{Cai2011}. 

Chen et al.\cite{Chen2017} also performed simulations without HI in their study. Rather than using LD, however, they destroyed HI within the MPCD framework by randomizing the solvent positions and velocities. Interestingly, in contrast with our findings, their simulations predict that the nanoparticle and polymer center-of-mass subdiffusive behavior decouples in the absence of HI. We hypothesize that this discrepancy arises because their approach for destroying HI does not preserve the long-time nanoparticle and polymer center-of-mass diffusive behavior, and thus it also likely distorts the relative time scales associated with other relaxation processes in the system. Hydrodynamic interactions influence various aspects of solution dynamics.  Unfortunately, there is no unique approach for removing HI from simulations that would allow an unambiguous characterization of its contributions at {\it all} time scales. The approach that we have adopted preserves the long-time particle and polymer relaxation time scales under dilute conditions. These relaxations are influenced by drag from the solvent, but, by definition, not by many-body contributions arising from momentum transfer between nanoparticles and polymers. Even though these additional contributions may influence relaxations at finite solute concentrations, this does not imply that they dictate the physical mechanisms controlling nanoparticle-polymer coupling. Indeed, our results demonstrate that when the long-time relaxations of the system are preserved, the nanoparticle and polymer center-of-mass subdiffusive behavior remains strongly coupled even in the absence of many-body nanoparticle-polymer hydrodynamic correlations.

\section{Conclusion}
We simulated the dynamics of nanoparticles in semidilute polymer solutions with and without long-range hydrodynamic interactions. The long-time nanoparticle dynamics were well described by recent theoretical predictions based on coupling to segmental relaxations; this coupling theory also captures the short-time dynamics for particles smaller than or much larger than the polymer correlation length. In agreement with experiments, however, the simulations revealed a surprisingly broad crossover regime where the subdiffusive exponent was larger than predicted. Analysis of the simulation trajectories suggests that the nanoparticles couple to the subdiffusive dynamics of the polymer center-of-mass on short time scales, which provides an additional mechanism by which nanoparticles can move through the solution. Analogous physical pictures have been proposed to explain tracer dynamics in colloidal glasses \cite{Sentjabrskaja2016} and crowded biological material,\cite{Weiss2004} in which the coupling between tracer and crowder dynamics leads to subdiffusive dynamics. 

\begin{acknowledgement}
We thank Dan Bolintineanu for useful discussions regarding the implementation of MPCD in LAMMPS. This work was supported by the Welch Foundation (Grants E-1882 (JCP) and E-1869 (JCC)) and the National Science Foundation (CBET-1705968, to JCP and JCC). Computational resources were generously provided by the Center for Advanced Computing and Data Systems at the University of Houston. AN acknowledges funding from the German Research Foundation (DFG) under project number NI 1487/2-1. This research (MPH) is part of the Blue Waters sustained-petascale computing project, which is supported by the National Science Foundation (awards OCI-0725070 and ACI-1238993) and the state of Illinois. Blue Waters is a joint effort of the University of Illinois at Urbana-Champaign and its National Center for Supercomputing Applications.
\end{acknowledgement}

\bibliography{biblio}

\begin{tocentry}
\includegraphics[width = 3.25 in]{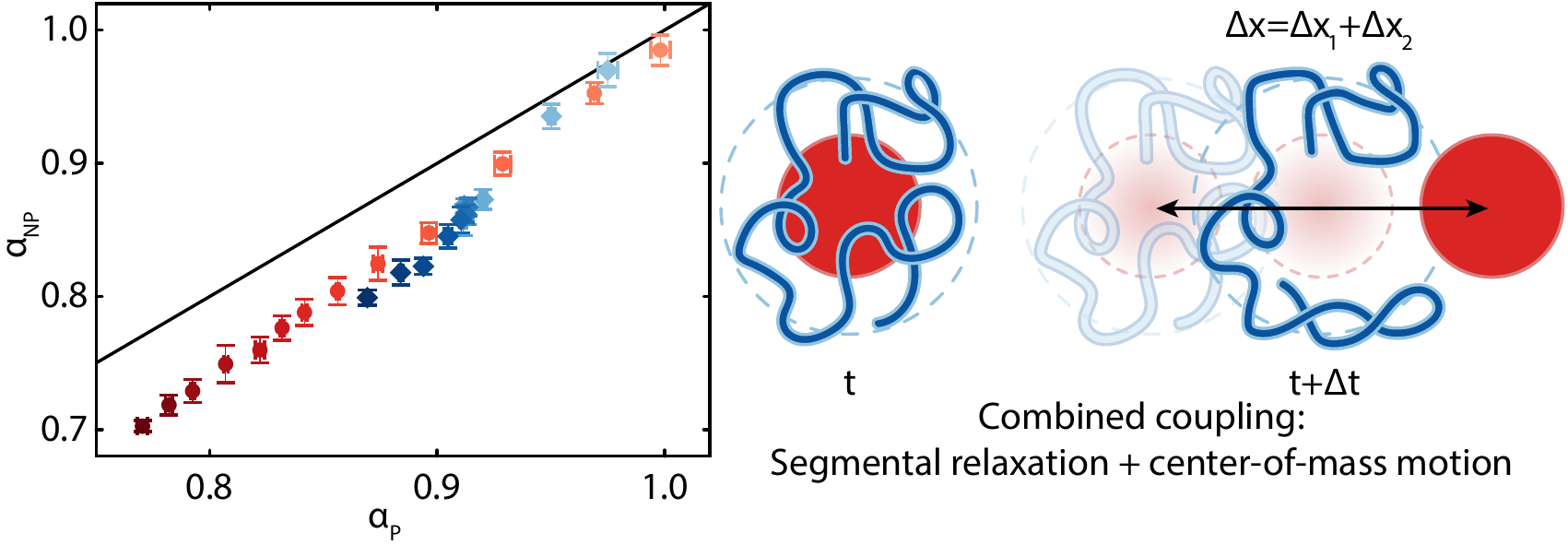}
\end{tocentry}

\end{document}